\documentclass[twocolumn,showpacs,preprintnumbers,amsmath,amssymb]{revtex4}

\usepackage{graphicx}% Include figure files
\usepackage{graphicx,epstopdf}
\usepackage{xcolor}
\usepackage{dcolumn}% Align table columns on decimal point
\usepackage{bm}% bold math
\usepackage{mathrsfs} % script-like, curvy letters.
\usepackage{mathtools}
\usepackage{hyperref}
\usepackage{bbm}
\begin{document}

%\preprint{APS/123-QED}
\title{Higher-order interactions promote chimera states}
\author{Srilena Kundu}
\author{Dibakar Ghosh}
\email{diba.ghosh@gmail.com}
\affiliation{Physics and Applied Mathematics Unit, Indian Statistical Institute, 203 B. T. Road, Kolkata-700108, India
}

\date{\today}

\begin{abstract}
Since the discovery of chimera states, the presence of a nonzero phase lag parameter turns out to be an essential attribute for the emergence of chimeras in a nonlocally coupled identical Kuramoto phase oscillators' network with pairwise interactions. In this letter, we report the emergence of chimeras without phase lag in nonlocally coupled identical Kuramoto network owing to the introduction of non-pairwise interactions. The influence of added nonlinearity in the coupled system dynamics in the form of simplicial complexes mitigates the requisite of a nonzero phase lag parameter for the emergence of chimera states. Chimera states stimulated by the reciprocity of the pairwise and non-pairwise interaction strengths and their multistable nature are characterized with appropriate measures and are demonstrated in the parameter spaces.
\end{abstract}


\maketitle

%\tableofcontents

\paragraph*{Introduction.}\label{intro}

Many natural and human-made systems \cite{newman2003structure, boccaletti2006complex} ranging from physics, biology, engineering and social sciences are modeled as networks whose constituents are represented as dynamical systems interacting among themselves through various links. The interplay of the network configuration and the underlying dynamical processes often gives rise to various nonlinear collective phenomena \cite{arenas2008synchronization, a2001synchronization, panaggio2015chimera, saxena2012amplitude} that has immense applicability across wide disciplines. So far, the connectivity among the dynamical elements of such complex system has been mostly described by interactions within a pair of nodes. However, recent progresses \cite{carletti2020dynamical, battiston2020networks} in complex system research have significantly highlighted the prominence of non-pairwise interactions in addition to pairwise interactions. In several real networks, such as brain network \cite{petri2014homological, sizemore2018cliques}, social network \cite{benson2016higher}, ecological interaction \cite{grilli2017higher}, random walk \cite{carletti2020random}, scientific collaboration network \cite{patania2017shape}, social contagion \cite{iacopini2019simplicial} etc., simple pairwise interactions may not be sufficient to unravel the prevailing physical mechanisms. The inherent dynamical processes are efficiently captured while taking into account higher-order interactions \cite{battiston2020networks, bianconi2021higher, torres2021and, battiston2021physics}, that have been widely adopted in the literatures, recently.

\emph{Simplicial complex} \cite{giusti2016two, salnikov2018simplicial, torres2020simplicial} formed by simplices of different dimensions, is one such topological framework that is often used to represent the underlying structural configuration of higher-order interaction networks. The interaction among $n+1$ dynamical units is represented as $n$ simplex, so a 2-simplex typifies three-body interaction, 3-simplex typifies four body interaction, etc. Very  recently, researchers took interest in studying the collective dynamical behavior \cite{tanaka2011multistable, skardal2019abrupt, millan2020explosive, lucas2020multiorder, gambuzza2021stability, skardal2020higher, ghorbanchian2021higher, zhang2021unified} observed in higher-order networks which reveals some exciting finding due to the incorporation of non-pairwise interactions. 

Synchronization \cite{arenas2008synchronization, a2001synchronization, boccaletti2018synchronization} is one such collective phenomenon where the interactions among the dynamical entities play a crucial role. Recent investigations reveal how the presence of higher-order interactions affects the transition scenario to synchronization and also triggers the emergence of various synchronization states. The Kuramoto phase oscillator \cite{kuramoto2003chemical, rodrigues2016kuramoto} is considered as the simplest model for describing the synchronization phenomena, which when generalized in a network setup with higher-order interactions, promotes the emergence of explosive synchronization \cite{skardal2019abrupt, millan2020explosive, skardal2020higher}, complete synchronization \cite{gambuzza2021stability}, cluster synchronization \cite{zhang2021unified}, etc. 

An intriguing collective dynamics where both the synchronization and desynchronization state coexists simultaneously is known as \emph{chimera state} \cite{kuramoto2002coexistence, abrams2004chimera}, that results from the symmetry breaking of the network. The emergence of this exotic state has been massively explored by considering diverse network topologies \cite{laing2009chimera, martens2010solvable, kundu2018chimera, panaggio2013chimera, omel2012stationary, kundu2019chimera, panaggio2015chimera1, kundu2021amplitude, makarov2019multiscale, zhu2014chimera, ghosh2016birth, majhi2017chimera, hizanidis2016chimera} and coupling configurations \cite{bera2017chimera, laing2015chimeras, bera2017coexisting, abrams2006chimera, hizanidis2014chimera, nkomo2013chimera, omel2008chimera, schmidt2014coexistence, sethia2014chimera, mishra2015chimeralike}, in the last two decades. However, these studies are concentrated only on networks having pairwise interaction. Very recently, Zhang et al. \cite{zhang2021unified}, as an application to their proposed unified theory for stability analysis of synchronization patterns, reported the emergence of chimera states in networks of optoelectronic oscillators in the presence of non-pairwise interactions. 

Till now, innumerable studies \cite{panaggio2015chimera, kuramoto2002coexistence, abrams2004chimera, abrams2006chimera, maistrenko2014cascades, laing2009dynamics} in networks with pairwise interactions have confirmed the crucial importance of a non-zero phase lag parameter, in order to observe chimera states in networks of nonlocally coupled identical Kuramoto phase oscillators. These fascinating chimera states are indeed possible without the presence of phase lag, however the network cannot be simply coupled non-locally with identical oscillators \cite{wang2011synchronization, frolov2020chimera}. Either some non-homogeneity should be introduced among the individual oscillators or the coupling configuration should differ from that of the usual non-local topology. On contrary to this, for the very first time to the best of our knowledge, in this letter we report the emergence of chimera states in nonlocally coupled identical Kuramoto phase oscillators in the absence of the crucial phase lag parameter, when the limit of pairwise interactions is removed. Specifically, in this study, we consider coupled oscillator simplicial complexes with nonlocal interaction, and explore the impact of 2-simplexes and 3-simplexes (see Supplemental Material \cite{sm}) on the advent of chimera states. Taking account of the higher-order interaction terms, our investigation unveils that chimera states can be observed extensively without pahse lag. In fact, we found that the chimera region broadens in the parameter space with the increase of the non-pairwise interaction strength. The influence of initial conditions have already been well established in the literature for the emergence of chimera states in pairwise interacting network \cite{martens2016basins, rakshit2017basin, dos2020basin, faghani2018effects}. Here, we observe that the variation of initial conditions in presence of higher-order inetractions promotes the coexistence of multiple states, which are characterized by quantifying the basin stability of those states. Further, we explore the most probable route of transition from incoherent to coherent dynamics as the 2-body interaction strength increases in the presence of higher-order interaction. 

\paragraph*{Nonlocally coupled higher-order Kuramoto network.}

To explore the effect of higher-order interaction on the emergence of chimera states, we consider a network of $N$ non-locally coupled identical phase oscillators with 3-body and 4-body interaction terms along with 2-body interactions. Specifically, we consider a simplicial complex of $N$ nodes having simplices of dimension $P = \{2, 3\}$, where $P$ is the non-local coupling radius. %The schematic diagram of the network is depicted in Fig. $\ref{fig0}$.
The mathematical equation of higher-order Kuramoto network without phase lag is given by 
\begin{equation}\label{eq1}
\begin{array}{lcl}
\dot{\theta_i} = \omega + \frac{\epsilon_1}{k_1}\sum\limits_{j=1}^N A_{ij}\sin(\theta_j-\theta_i) \\ ~~~~+ \frac{\epsilon_2}{k_2}\sum\limits_{j=1}^N \sum\limits_{k=1}^N B_{ijk} \sin(\theta_j + \theta_k - 2\theta_i) \\ ~~~~+ \frac{\epsilon_3}{k_3}\sum\limits_{j=1}^N \sum\limits_{k=1}^N \sum\limits_{l=1}^N C_{ijkl} \sin(\theta_j + \theta_k + \theta_l - 3\theta_i),
\end{array}
\end{equation}
where $\theta_i$ denotes the phase of the oscillator placed at position $x_i$ and $\omega$ is the identical natural frequency of the oscillators. The underlying network topology is considered to be non-local having $P$ pairwise interactions on each side of each of the $N$ nodes. The 1-, 2- and 3-simplex interactions are encoded in the adjacency matrix $A$, and the adjacency tensors $B, C$, respectively, such that $A_{ij}=1$ if there is a link $(i,j)$ ($0$ otherwise), $B_{ijk}=1$ if there is a triangle $(i,j,k)$ ($0$ otherwise), $C_{ijkl}=1$ if there is a tetrahedron $(i,j,k,l)$ ($0$ otherwise). The parameter $\epsilon_q$ represents the strength of the $q$-simplex interaction and $k_q$ is the $q$-simplex degree of each node across the network for $1 \leq q \leq P$. For simplicity, here we assume $\epsilon'_q = \epsilon_q/k_q$. The higher-order sinusoidal coupling functions associated with each node $i$ are chosen in such a way that they remain symmetric with respect to $i$. 

In the following, we explore the consequences of introducing higher-order interactions in the classical nonlocally coupled Kuramoto phase oscillator network without any phase lag. In this study, we consider the network size $N=100$ and the natural frequency of oscillators $\omega=1$. All the numerical results are implemented using fifth order Runge-Kutta-Fehlberg integration scheme with fixed time step $dt = 0.01$. We choose the boundary conditions to be periodic and the initial conditions $\theta_i(0)$ are chosen randomly \cite{random} from the interval $[0,~2\pi]$ that are fixed for all the simulations throughout the letter (unless otherwise mentioned).

\begin{figure}[ht]
	\centerline{
	\includegraphics[scale=0.35]{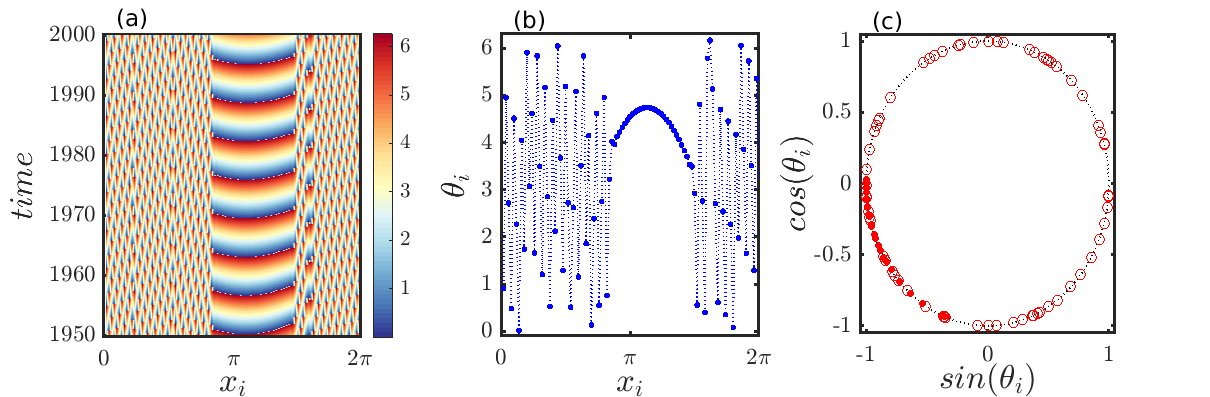}}
	\caption{(a) Spatiotemporal evolution of the observed chimera state in the absence of phase lag in a non-locally coupled higher-order Kuramoto phase oscillator network with 2-simplex interaction. The colorbar corresponds to the state of the phase variable $\theta_i$ of the oscillator at position $x_i$. (b) Snapshot of the phases $\theta_i$ at position $x_i$ depicting the coherent and incoherent subpopulations at the final time $t=2000$, (c) the distribution of the phases over the complex unit circle, where the synchronized cluster is identified with red dots and the incoherent cluster is associated with the markers plotted in red circles. The parameter values are fixed at $N = 100, P=2, \omega =1, \epsilon'_1 = 0.1$, and $\epsilon'_2 = 0.3$.}
	\label{fig1}
\end{figure}

\paragraph*{Results.}
First, we bring out the analysis by considering $P=2$, that means only 2-simplex interactions are present in this network along with 1-simplex interaction. Therefore, in Eq. \eqref{eq1}, the last term associated with the 3-simplex interaction will remain absent in this case. Figure \ref{fig1} unfolds the emergence of coexisting synchronized and desynchronized dynamics in presence of non-zero 2-simplex interaction strength $\epsilon'_2 = 0.3$ when the 1-simplex interaction strength $\epsilon'_1 = 0.1$ is considered. The spatiotemporal pattern of the observed chimera state for the chosen parameter values is depicted in Fig. \ref{fig1}(a). At a particular time instant $t = 2000$, the phase distribution of the oscillators is shown in Fig. \ref{fig1}(b). These two subfigures clearly demonstrate the emergence of a single coherent domain in between two incoherent sub-populations. This chimera state is further illustrated in Fig. \ref{fig1}(c), where individual oscillators belonging to the synchronized cluster are depicted with red dots and the oscillators plotted with red circles belong to the incoherent sub-population that are distributed randomly over the unit circle.

\begin{figure}[ht]
	\centerline{
	\includegraphics[scale=0.32]{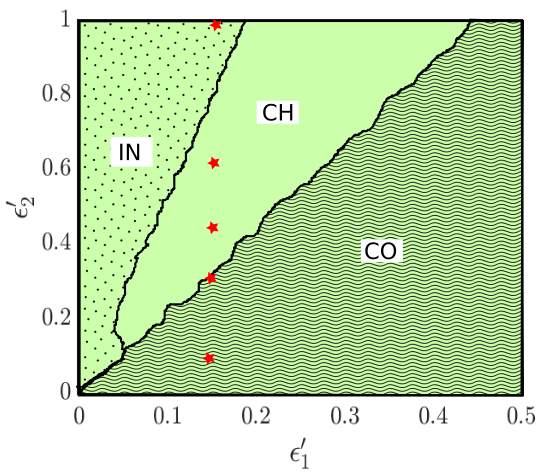}}
	\caption{Characterization of three different dynamical states, namely incoherent (IN), chimera (CH), and coherent (CO) states as an interplay of the 1-simplex and 2-simplex interaction strengths $\epsilon'_1$ and $\epsilon'_2$, respectively. The states are respectively classified depending on the values of $\mbox{SI}=1, \mbox{SI}\in (0,1)$, and $\mbox{SI}=0$, as computed from Eq. \eqref{eq2}. The following parameter values are used to generate the figure: $\beta=20, \delta=0.5$. The time average $\langle \cdot \rangle_t$ in Eq. \eqref{eq3} is taken over  $2\times10^3$ time iterations.}
	\label{fig2}
\end{figure}

In order to properly distinguish the three different parametric regimes corresponding to the three different dynamical states, namely coherent, chimera, and incoherent, we adopt the conventional statistical measure  strength of incoherence (SI) \cite{gopal2014observation}, which is calculated using the following formula
\begin{equation}\label{eq2}
\begin{array}{lcl}
\mbox{SI} = 1 - \frac{\sum_{m=1}^\eta s_m}{\eta}, ~~~~s_m = \Theta[\delta-\sigma(m)],
\end{array}
\end{equation}
where $\Theta(\cdot)$ is the Heaviside step function and $\delta$ is a predefined threshold. The oscillators are subdivided into $\eta$ bins of equal length $\beta = \frac{N}{\eta}$, and local standard deviation $\sigma(m)$ is measured in each of these bins as
\begin{equation}\label{eq3}
\sigma(m) = \Bigg\langle \sqrt{\frac{1}{\beta} \sum\limits_{i=\beta(m-1)+1}^{m\beta} (\Delta\theta_i - \langle \Delta \theta \rangle)^2} \Bigg\rangle_t, m=1, 2, ..., \eta.
\end{equation}
Here, the variable $\Delta \theta_i = \theta_{i+1} - \theta_i$ represents the difference of phases between two adjacent oscillators for $i=1,2, ..., N$; and $\langle \Delta \theta \rangle = \frac{1}{N} \sum_{i=1}^N \Delta \theta_i$. The interplay of the 1-simplex and 2-simplex interaction strengths are portrayed in Fig. \ref{fig2} in the $\epsilon'_1-\epsilon'_2$ parameter space using the SI measure as described in Eq.~\eqref{eq2} for the initial condition discussed in \cite{random}. The incoherent (IN), coherent (CO), and chimera (CH) regions in the parameter space are characterized by the values $\mbox{SI}\simeq 1, \mbox{SI}=0$, and $\mbox{SI} \in (0,1)$, respectively. In the absence of phase lag
\begin{figure}[ht]
	\centerline{
	\includegraphics[scale=0.38]{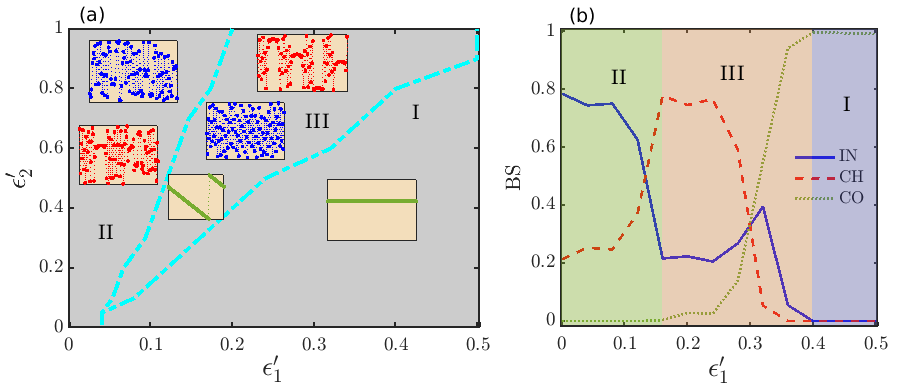}}
	\caption{(a) Demonstration of the various stability regions in the $\epsilon'_1-\epsilon'_2$ parameter space characterized by the quantification of the basin stability measure. Region I: existence of only coherent dynamics, Region II: coexistence of incoherent and chimera dynamics, Region III: coexistence of chimera, incoherent and coherent dynamics. The inset figures in the respective regions elucidate the snapshots of the phases for different choices of initial conditions. (b) Variation of the basin stability $\mbox{BS}$ is instantiated by varying the 1-simplex coupling strength $\epsilon'_1$ for a particular choice of 2-simplex interaction strength $\epsilon'_2 = 0.7$. The probability of obtaining three different dynamical states is shown in three different colors. Blue (solid), red (dashed) and green (dotted) colors correspond to the incoherent, chimera, and coherent dynamics, respectively.}
	\label{fig3}
\end{figure}
 in pairwise coupled network, 1-simplex coupling strength gives rise to the synchronization phenomena \cite{panaggio2015chimera}. This occurrence is also verified in this figure for $\epsilon'_2=0$, which produces only the coherent region. For very smaller values of the 2-simplex interaction strength $\epsilon'_2 > 0$, a direct transition from incoherent to coherent dynamics is noticed in the parameter space as $\epsilon'_1$ increases. However, the chimera region shows up beyond a certain value of $\epsilon'_2$ and this region keeps enlarging with higher $\epsilon'_2$. Hence, simply the addition of 2-simplex interaction in the networked equation significantly promotes the chimera dynamics which was previously impossible to achieve in the absence of phase lag in a nonlocally coupled network of identical phase oscillators with 1-simplex interaction. However, our thorough numerical analysis uncovers the possibility of obtaining multiple stable dynamics depending on the initial phase values at any particular choice of the coupling pair ($\epsilon'_1, ~\epsilon'_2$). Particularly, we have found that the chimera dynamics in the $(\epsilon'_1, ~\epsilon'_2)$ parameter space coexists either only with the incoherent dynamics or both the incoherent and coherent dynamics. Thus the considered higher-order Kuramoto network exhibits multistable phenomena, which is in congruence with the previously observed multistable behavior \cite{tanaka2011multistable, skardal2019abrupt} due to the inclusion of non-pairwise interactions in the network. To characterize the coexistence of multiple stable dynamical states depending on the choice of the initial values, we stick to the basin stability measure \cite{rakshit2017basin} that quantifies the volume of the basin leading to a particular dynamical state. We choose $V_0 = 1000$ random initial conditions $\theta_i(0)$ uniformly distributed over the interval $[0, ~2\pi]$ for each of the $i$th oscillator, to quantify the fraction of initial conditions leading to either of incoherent, chimera or coherent state. Thus the basin stability is measured according to the formula $\mbox{BS} = \frac{V_s}{V_0}$, such that $\mbox{BS} \in [0,~1]$, where $V_s$ is the number of initial conditions leading to any particular state. The values $\mbox{BS}=0$ and $\mbox{BS}=1$, respectively, correspond to the instability and monostability of any particular state, whereas the coexistence of multiple states are indicated by the values $0<\mbox{BS}<1$. Figure \ref{fig3}(a) illustrates the occurrence of three different stability regions characterized by means of the $\mbox{BS}$ measure in the $\epsilon'_1-\epsilon'_2$ parameter space that are demarcated by the cyan boundaries. Region I corresponds to the monostable region depicting the presence of only the coherent dynamics, region II exhibits the bistable region where the coexistence of incoherent and chimera state is detected, and region III is the region of tristability where all the three dynamical states coexist for specific choices of initial conditions. Here, the increased nonlinearity due to the 2-simplex interactions dominates the basin of the coherent state over the basin of the incoherent and/or chimera states. As a result, the basin of the incoherent and/or chimera state becomes much larger compared to the basin of the coherent state in regions II and III. Even though the coherent state is always linearly stable for identical Kuramoto oscillators \cite{lucas2020multiorder}, the significant contribution of the linear approximations owing to the higher-order interactions in the linearized system fails to easily detect the coherent state in the parameter region II. That is why we observe only the emergence of incoherence and chimera in region II for 1000 realizations over the initial conditions from the basin $[0, 2\pi]$. However, significantly increasing the number of realizations or reducing the basin of initial conditions, the coherent state can be observed in region II. How the variation of the initial conditions affects the emergence of different states by varying the 1-simplex interaction strength $\epsilon'_1$, is delineated in Fig. \ref{fig3}(b) for an exemplary value of the 2-simplex interaction strength $\epsilon'_2 = 0.7$. Blue (solid), red (dashed) and green (dotted) colors correspond to the basin stability of incoherent, chimera and coherent states, respectively. The figure substantiates that the probability of getting incoherent state is higher in region II than in region III, whereas in region III the chimera states are most probable and coherent states are very less probable. Also, the presence of only coherent state in region I is evidenced from the figure.

\begin{figure}[ht]
	\centerline{
	\includegraphics[scale=0.36]{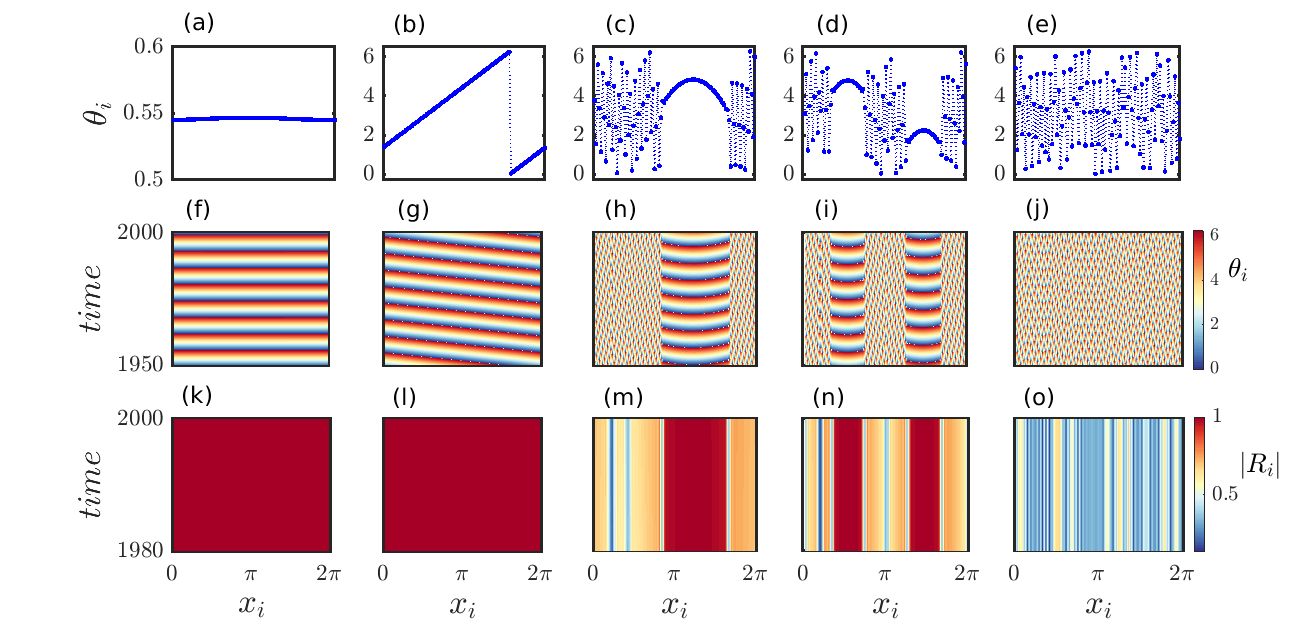}}
	\caption{Transition from coherent to incoherent dynamics (from left to right) when the 2-simplex interaction strength $\epsilon'_2$ is varied for a particular choice of 1-simplex interaction strength $\epsilon'_1 = 0.15$. Five different spatiotemporal patterns are manifested for values $\epsilon'_2 = 0.1, 0.3, 0.45, 0.62,$ and $1.0$, respectively, that correspond to the five red markers shown in Fig. \ref{fig2}. For lower values of $\epsilon'_2$, the dynamics remains coherent (a, f, k), which then shifts to coherent traveling wave when $\epsilon'_2$ is increased slightly (b, g, l). Further increment induces the chimera  dynamics (c, h, m) which then switches to multichimera state (d, i, n) and then finally to incoherent state (e, j, o) as $\epsilon'_2$ advances. Snapshots of the phase variables $\theta_i$ are depicted in the upper row at $t=2000$ and their corresponding spatiotemporal behavior is displayed in the middle row. The bottom row shows the spatiotemporal variation of the order parameter amplitude $|R_i|$ for the respective values of $\epsilon'_2$, such that $|R_i|=1$ for the coherent dynamics and $|R_i| < 1$ for the incoherent dynamics.}
	\label{fig4}
\end{figure}

Besides, we also investigate the transition route from synchronized to desynchronized dynamics as $\epsilon'_2$ increases for a particular value of $\epsilon'_1$. Five qualitatively different dynamical states are detected during our inspection for our chosen initial conditions, the dynamics of which are plotted in Fig. \ref{fig4} for five different values of $\epsilon'_2$ (corresponding to the five red markers shown in Fig. \ref{fig2}). For fixed $\epsilon'_1 = 0.15$, coherent dynamics is observed for smaller values of $\epsilon'_2$, a typical snapshot of which is depicted in Fig. \ref{fig4}(a) taking $\epsilon'_2 = 0.1$. As $\epsilon'_2$ increases, the dynamics changes to coherent traveling wave. An exemplary snapshot is portrayed in Fig. \ref{fig4}(b) for $\epsilon'_2 = 0.3$. Further increment in $\epsilon'_2$ shifts the dynamics from coherent to chimera state with single coherent cluster, as shown in Fig. \ref{fig4}(c) for $\epsilon'_2 = 0.45$. With the increase of 2-simplex interaction strength the coherent region is subdivided into more than one cluster and induces the multichimera state. The corresponding dynamics is shown in Fig. \ref{fig4}(d) for $\epsilon'_2 = 0.62$. Finally, beyond a certain value of $\epsilon'_2$, incoherent dynamics is triggered in the system, whose snapshot is illustrated in Fig. \ref{fig4}(e) for $\epsilon'_2 = 1.0$. Spatiotemporal behavior of all these different dynamical states are demonstrated in the middle row that manifest their stationarity. To determine the level of coherence among the neighboring oscillators, we define a complex order parameter $R_i$ for each of the $i$th node as $R_i = (R_{1_i}+R_{2_i})/2$, where $R_{q_i}$ for $q = 1, 2$ gives an essence of the coherence among the oscillators due to the $q$-simplex interaction which is defined for the $i$th oscillator as $R_{1_i} = \frac{1}{k_1} \sum\limits_{j=1}^N A_{ij} e^{\mathbbm{i}\theta_j} $ and $R_{2_i} = \frac{1}{k_2} \sum\limits_{j=1}^N \sum\limits_{k=1}^N B_{ijk} e^{\mathbbm{i}(\theta_j + \theta_k)} $, where $\mathbbm{i} = \sqrt{-1}$. The spatiotemporal variation of amplitude $|R_i|$ of oscillator at $x_i$ is depicted in the bottom row of Fig. \ref{fig4} for the corresponding values of $\epsilon'_2$, which confirms that the synchronized cluster takes value $|R_i| = 1$, while $|R_i| < 1$ for the incoherent cluster.

\begin{figure}[ht]
	\centerline{
	\includegraphics[scale=0.5]{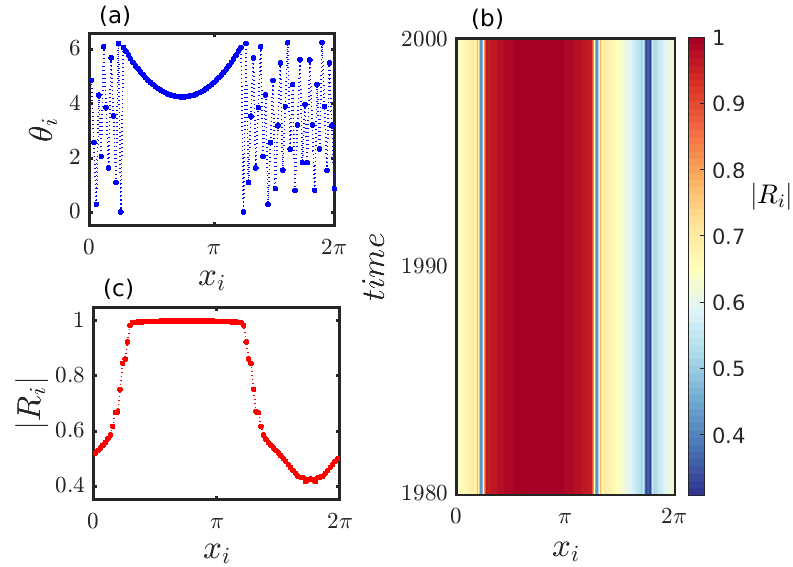}}
	\caption{(a) Snapshot of the chimera state for the pairwise and triangular interaction strengths $\epsilon'_1 =0.3, \epsilon'_2=0.8$, respectively. (b) Spatiotemporal variation of the order parameter amplitude $|R_i|$ computed directly from the numerical simulations, and (c) profile of $|R_i|$ at a particular time when computed using Eqs. \eqref{eq6}-\eqref{eq7} obtained from the Ott-Antonsen approach.}
	\label{fig5}
\end{figure}

\begin{figure}[ht]
	\centerline{
	\includegraphics[scale=0.35]{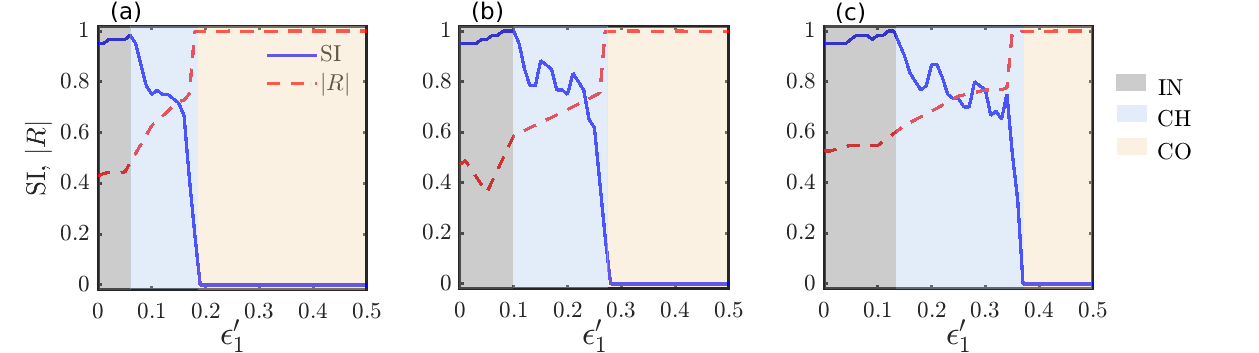}}
	\caption{Characterization of the incoherent (IN), chimera (CH) and coherent (CO) region depending on the values of $\mbox{SI} \simeq 1, 0<\mbox{SI}<1, \mbox{SI} = 0$, respectively, when the 1-simplex interaction strength $\epsilon'_1$ is varied. The transition is also validated by plotting the theoretically derived $|R|$ values. The transition point at which $\mbox{SI}=0$ coincides with the point at which $|R|=1$, which determines the onset of coherent dynamics. Three body interaction strength $\epsilon'_2$ is fixed at (a) $\epsilon'_2 = 0.4$, (b) $\epsilon'_2 = 0.6$, and (c) $\epsilon'_2 = 0.8$.}
	\label{fig6}
\end{figure}

We look for some theoretical insights into the observed chimera dynamics by using the Ott-Antonsen \cite{ott2008low, laing2009dynamics} approach in the thermodynamic limit $N \to \infty$. Although, the OA reduction is more widely applicable to systems of oscillators with nonidentical frequencies, this method can also be utilized effectively for homogeneous networks \cite{marvel2009identical, hong2011conformists, xu2016collective}. Using the complex order parameters $R_{1_i}$ and $R_{2_i}$, Eq. \eqref{eq1} can be rewritten as
\begin{equation}\label{eq4}
\dot{\theta_i} = \omega + \frac{1}{2\mathbbm{i}}[H_i e^{-\mathbbm{i}\theta_i} - {H_i}^* e^{\mathbbm{i}\theta_i}],
\end{equation}
where $H_i = \epsilon_1 {R_1}_i + \epsilon_2 {R_2}_i e^{-\mathbbm{i}\theta_i}$ and $^*$ denotes the complex conjugate. In the continuum limit, the state of the system can be represented by a probability density function $f(\theta, t)$, that gives the fraction of oscillators with phases lying between $\theta$ and $\theta + d\theta$ at time $t$. Since the number of oscillators in the system is conserved, $f$ satisfies the continuity equation

\begin{equation}\label{eq5}
\frac{\partial f}{\partial t} + \frac{\partial}{\partial \theta} (fv) = 0.
\end{equation}
Here $v = \dot{\theta}$ is given in Eq. \eqref{eq4} and $f$ can be expanded in a  Fourier series of the form $f(\theta,t) = \frac{1}{2\pi} [1+\sum\limits_{n=1}^\infty (\{h(x,t)\}^n e^{\mathbbm{i}n\theta} + \{h^*(x,t)\}^n e^{-\mathbbm{i}n\theta})]$; $\{h(x,t)\}^n$ being the $n$th Fourier coefficient. Substituting $f$ into Eq. \eqref{eq5} and comparing the coefficient of the term $e^{\mathbbm{i}n\theta}$, we obtain the time evolution of the variable $h(x_i,t)$ associated with the oscillator at position $x_i$ as
\begin{equation}\label{eq6}
\begin{array}{lcl}
\frac{\partial h (x_i)}{\partial t} = -\mathbbm{i}\omega h(x_i) + \frac{1}{2} \Big[\epsilon_1\Big({R_1}_i^*-{R_1}_ih(x_i)^2\Big) \\~~~~~~~~~~~+ \epsilon_2\Big({R_2}_i^*h(x_i)^{-1}-{R_2}_ih(x_i)^3\Big)\Big].
\end{array}
\end{equation}
Also, the order parameters $R_{1_i}, R_{2_i}$ can be derived in terms of $h(x_i,t)$ as
\begin{equation}\label{eq7}
\begin{array}{lcl}
R_{1_i} = \frac{1}{k_1}\sum\limits_{j=1}^N A_{ij} e^{\mathbbm{i}\theta_j} = \frac{1}{k_1}\sum\limits_{j=1}^N A_{ij} \int\limits_0^{2\pi}f(\theta_j,t) e^{\mathbbm{i}\theta_j} d\theta_j \\~~~~~~~~~~~~~~~~~~~~~~~~~~= \frac{1}{k_1} \sum\limits_{j=1}^N A_{ij}{h^*(x_j)}, \\
R_{2_i} = \frac{1}{k_2}\sum\limits_{j=1}^N \sum\limits_{k=1}^N B_{ijk} e^{\mathbbm{i}(\theta_j+\theta_k)} \\~~~~~= \frac{1}{k_2} \sum\limits_{j=1}^N \sum\limits_{k=1}^N B_{ijk} \int\limits_0^{2\pi} f(\theta_j,t) e^{\mathbbm{i}\theta_j} d\theta_j \int\limits_0^{2\pi} f(\theta_k,t) e^{\mathbbm{i}\theta_k} d\theta_k \\~~~~~= \frac{1}{k_2} \sum\limits_{j=1}^N \sum\limits_{k=1}^N B_{ijk} h^*(x_j) h^*(x_k).
\end{array}
\end{equation}
The evolution of the variable $h(x_i,t)$ as well as the order parameters $R_{q_i}$ for $q=1, 2$, can be determined by using Eqs. \eqref{eq6} and \eqref{eq7}, simultaneously. While solving these equations, the order parameter values $R_i$ derived from the phase values obtained from direct numerical simulation of Eq. \eqref{eq1}, are considered as the initial conditions $h(x_i,0)$. We compute the average $R_i$ from Eq. \ref{eq7} and plot their amplitude in Fig. \ref{fig5}(c) for a particular choice of the interaction strengths $(\epsilon'_1, \epsilon'_2) = (0.3, 0.8)$. The corresponding profile of the phases at a particular time $t=2000$ is depicted in Fig. \ref{fig5}(a) and the evolution of the order parameter amplitude $|R_i|$ computed directly from the numerical
 simulations is illustrated in Fig. \ref{fig5}(b). This figure substantiates the validation of the higher-order dynamics through the theoretically predicted order parameter values that are in well agreement with the numerically simulated dynamics of the system at least for a particular choice of the interaction strengths. In addition, the transition phenomena among the various dynamical states is exhibited in Fig. \ref{fig6} on the basis of SI measure and theoretically derived values of order parameter $|R|$ (average over the $|R_i|$ values for $i=1, 2, ..., N$) for three different values of the 2-simplex interaction strength $\epsilon'_2 = 0.4, 0.6$, and $0.8$, respectively. Three distinct regions corresponding to incoherent, chimera and coherent states are classified depending on the values of SI. For smaller values of $\epsilon'_1$ where $\mbox{SI}\simeq 1$, the dynamics is incoherent and order parameter $|R|$ takes lower values which then takes moderate values when the dynamics shifts to the chimera region. Finally, the transition point to coherent region where $|R| = 1$ is in well accordance with the transition point where $\mbox{SI}=0$.    

\paragraph*{Conclusion.}
Until now, chimera states have been investigated extensively from the perspective of networks where the dynamics is associated to the nodes and the interactions among the dynamical units are represented only by the links joining a pair of nodes. Previous research in this context, confirms the necessity of an additional phase lag parameter to develop chimera pattern in a nonlocally coupled network of identical Kuramoto phase oscillators. Presently, the rising interest in exploring various synchronization phenomena considering networks with non-pairwise interactions fosters the idea of investigating the emergence of fascinating chimera states in a network with higher-order interactions. In this letter, we discover the unusual occurrence of coexisting synchronous and asynchronous dynamics in the absence of phase lag in a nonlocally coupled identical Kuramoto network incorporating higher dimensional interactions. Specifically, we adopted the simplicial complex network topology and scrutinized the impact of higher-order simplexes on the emergence of distinct collective states like synchronization, desynchronization and chimera states. Considering $P=2$, we rigorously analyzed the parameter space containing 1-simplex and 2-simplex interaction strengths and explained the possible route of transition from coherent to incoherent dynamics when the 2-simplex interaction strength increases for a particular choice of 1-simplex interaction strength. We found that  the inclusion of higher-order terms gives rise to the multistable behavior where either two or three of the incoherent, chimera and coherent dynamics coexist. The regions of monostability, bistability and tristability are characterized by calculating the basin stability measure. Moreover, we utilized the Ott-Antonsen approach in the large $N$ limit and derived the evolution equation of the order parameter, which are found to be in good agreement when compared with their numerically computed counterpart. Besides, we also analyzed the network for $P=3$ (see Supplemental Material \cite{sm}) with the addition of 3-simplex interactions and demonstrated how the inclusion of four body interactions affects the observed phenomena in the parameter space.

\begin{figure}[ht]
	\centerline{
		\includegraphics[scale=0.4]{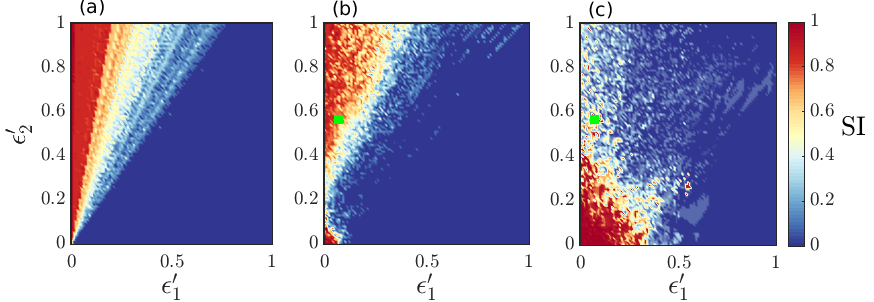}}
	\caption{Characterization of the incoherent, chimera and coherent dynamics in the $\epsilon'_1-\epsilon'_2$ parameter space for the nonlocal coupling range $P=3$. Three different values of the 3-simplex interaction strength (a) $\epsilon'_3=0.0$, (b) $\epsilon'_3=0.2$, and (c) $\epsilon'_3=1.0$ are considered. The colorbar corresponds to the $\mbox{SI}$ values that helps in identifying the incoherent ($\mbox{SI} \simeq 1$) and coherent ($\mbox{SI}=0$) region depicted with dark red and dark blue color, respectively. Intermediate colors correspond to the chimera region for which $0<\mbox{SI}<1$. Other parameter values are same as in Fig.\ 2 of the main text.}
	\label{figSM1}
\end{figure}

\begin{figure}[ht]
	\centerline{
		\includegraphics[scale=0.52]{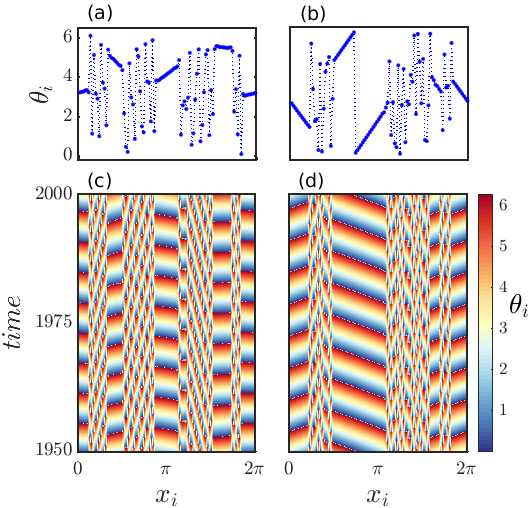}}
	\caption{Illustration of the chimera dynamics for $P=3$ due to the introduction of four body interactions in the network. Simultaneous interplay of all the 1-, 2- and 3-simplex interaction strengths gives rise to the chimera state with traveling coherent part(s). The interaction strengths are $\epsilon'_1 = 0.07, \epsilon'_2 = 0.57$ as shown in Fig. \ref{figSM1} with green marker points. The 3-simplex interaction strength $\epsilon'_3$ is chosen as (a, c) $\epsilon'_3=0.2$, (b, d) $\epsilon'_3=1.0$.}
	\label{figSM2}
\end{figure}

\paragraph*{Supplemental Material: Higher order interactions promote chimera states}
Here, we analyze the nonlocally coupled network for $P=3$ and very briefly summarize the major outcomes in Figs. \ref{figSM1} and \ref{figSM2}, respectively. Apart from 1-simplex and 2-simplex interactions, here 3-simplex interactions (i.e., four body interaction) are also taken into consideration. So, we mostly investigate the impact of the additional 3-simplex interaction strength $\epsilon'_3$ on the occurrence of the chimera states in presence of both the 1-simplex and 2-simplex interaction strengths $\epsilon'_1$ and $\epsilon'_2$, respectively. In Fig. \ref{figSM1}, we delineate the evolution of the different dynamical regimes in the parameter space with the increase of $\epsilon'_3$, when both $\epsilon'_1$ and $\epsilon'_2$ are varied simultaneously. Using the $\mbox{SI}$ measure as discussed in Eqs.\ (2)-(3) of the main text, three different dynamical regions associated with the incoherent, chimera and coherent states can be characterized from the figure. We choose three distinct values of the 3-simplex interaction strength $\epsilon'_3 = 0.0, 0.2, 1.0$ to portray the variation in the $\epsilon'_1-\epsilon'_2$ parameter regions. The dark red and dark blue regions correspond to the incoherent and coherent dynamics, respectively, whereas the intermediate colors are associated with the chimera dynamics. The parameter space in the left panel for $\epsilon'_3=0.0$ is qualitatively similar to the one illustrated in Fig.\ 2 of the main text for the case when $P=2$. This means that increasing the nonlocal coupling range $P$ does not alter the dependence of the 1-simplex and 2-simplex interaction strengths on the emergence of the distinct collective phenomena. As soon as $\epsilon'_3$ is considered to be non-zero, a new incoherent region starts developing on the lower left corner of the parameter space and the chimera region appears on the boundary of the incoherent and the coherent region. As an exemplary demonstration of the chimera pattern observed in presence of the four body interaction strength $\epsilon'_3$, we plot the snapshots and their spatiotemporal evolution for $\epsilon'_2 = 0.2$ and $1.0$ in Fig. \ref{figSM2} corresponding to the values plotted with green markers in Fig. \ref{figSM1}. Here the key observation is that unlike the chimeras observed for $P=2$, in presence of 3-simplex interactions the coherent parts of the chimera state can exhibit traveling wave dynamics.

\par {\bf Acknowledgments:}
The authors would like to thank Md. Sayeed Anwar for helpful discussions.

%\section*{References}
%\bibliographystyle{iopart-num}
\bibliographystyle{apsrev4-1}

%\bibliography{ref} % References file

\begin{thebibliography}{72}
\bibitem [{ran()}]{random}%
	\BibitemOpen
	\href@noop {} {\bibinfo  {journal} {The random numbers are generated using
			rand() function in Fortran for a fixed seed = 123456}\ }\BibitemShut
	{NoStop}%
\end{thebibliography}

\end{document}